\documentclass[12pt,a4paper]{article}
\usepackage[T1]{fontenc}
\usepackage[utf8]{inputenc}
\usepackage{amsmath,amssymb}
\usepackage[english,russian]{babel}
\usepackage{graphicx}
\usepackage{titlesec}

\newcommand{\be}{\begin{equation}\label}
\newcommand{\ee}{\end{equation}}
\newcommand{\prt}{\partial}
\newcommand{\p}{\prime}

\begin{document}
	
	\selectlanguage{English}
	
	\title{\hrule\vspace{0.1cm} {\small \vspace{0.1cm}}\hrule \vspace{2.0cm}
		{The hydrogen atom:\\ 
			consideration  of the electron self-field	}}

\author{ 
	Biguaa L.V~\footnote{E-mail:~leon.006w@yandex.ru}\\
	Kassandrov V.V~\footnote{E-mail:~vkassan@sci.pfu.edu.ru}\\
	{$^{*}$
	\small \em Quantum technology center}\\ {\small \em Moscow State University}\\
	{$^{\dagger}$ \small \em Institute of Gravitation and Cosmology}, \\ {\small \em Peoples' Friendship University of Russia}
}
		\maketitle

	Keywords: {\it Spinor electrodynamics. Dirac-Maxwell system of equations. Solitonlike solutions. Spectrum of characteristics. ``Bohrian'' law for binding energies}

	\vspace{0.4cm}
	\noindent
PACS: 03.50.-z; 03.65.-w; 03.65.-Pm

\selectlanguage{Russian}

\section{THE HYDROGEN ATOM AND CLASSICAL FIELD MODELS OF ELEMENTARY PARTICLES}

It is well known that the successful description of
the hydrogen atom, first in Bohr’s theory, and then via
solutions to the Schrödinger equation was one of the
main motivations for the development of quantum
theory. Later, Dirac described almost perfectly the
observed hydrogen spectrum using his relativistic generalization of the Schrödinger equation. A slight deviation from experiment (the Lamb shift between $2s_{1/2}$  and $2p_{1/2}$ levels in the hydrogen atom) discovered later
was interpreted as the effect of electron interaction
with vacuum fluctuations of electromagnetic field and
explained in the framework of quantum electrodynamics (QED) using second quantization. 

Nonetheless, the problem of the description of the
hydrogen atom still attracts attention of researchers,
and, in some sense, is a ``touchstone'' for many new
theoretical developments. In particular, it is possible
to obtain the energy spectrum via purely algebraic
methods ~\cite{Zaytsev}. Theories of the ''supersymmetric''
hydrogen atom are also widespread~\cite{Wipf}.  In the framework of the classical ideas, the spectrum can be
explained by the absence of radiation for the electron motion along certain “stable” orbits.   This effect can be
related with the balance of radiation and  {\it absorption}  of
energy from a hypothetical random electromagnetic
background in the framework of {\it  stochastic electrodynamics}~\cite{Stohast}, or it takes place in soliton models of the
hydrogen atom~\cite{Rybakov}.

The difficulties related to the description of the
hydrogen atom in canonical relativistic quantum
mechanics and quantum electrodynamics (QED)  include the “strange” singularity $\sim r^{-\alpha},~\alpha = e^2/\hbar c \approx 0.007$ of the Dirac wave function at zero for s- and p- states
which appears in spite of the initial requirement of
solution regularity. Consideration of the magnetic
field of the proton via perturbation theory alone does
not seem quite correct since the magnetic potential
near zero is of the order $1/r^2$ and dominates over the
Coulomb potential. This problem was discussed, in
particular, in ~\cite{Vigier} (see also~\cite{Samson}),  where the possibility of
constructing a magnetic theory of nuclear forces was
considered.

Of special topicality is consideration of the electromagnetic self-field of the electron. From the point of
view of quantum ideas, the electron in the hydrogen
atom represents a spatially distributed system  according to the probability density $\Psi^+ \Psi$ whose elements
interact both with the field of the nucleus and between
each other. Since the absolute value of the electron
charge is equal to the proton charge, the proton field
turns out to be screened at large distances.

It is known that consideration of the self-consistent
collective field of electrons is necessary for {\it  multielectron }  atoms, and is phenomenologically implemented,
for example, in the Hartree–Fock method. The multielectron wave function, however, does not differ fundamentally from the single-electron function. Consideration of the self-action of the spatially distributed
electron charge density does not contradict the ideas
of its “point-like” character (see discussion in~\cite{Vlasov,Pitaev}).    

Note also that a similar effect of gravitational selfaction of a “free” quantum particle was proposed in~\cite{Diosi,Penrose,Tod}  and other works in the context of the hypothesis on the objective character of the process of reduction of the particle wave function.  

It should be noted that the problem of consideration of the electron self-field (self-energy) and its
effect on the observed energy levels is one of the central problems in the context of QED (see, e.g.,~\cite{Gaytler}). According to the most widely accepted ideas, the
effect is related to interaction of the electron with the
second quantized electromagnetic field of “vacuum
fluctuations”~ \cite{Akhiezer}.The calculation of its value via perturbation theory (see, e.g., ~\cite{Mohr}) results in infinite
shifts of energy levels which, after mass renormalization, are reduced to the values of radiative corrections
(Lamb shift etc.) that agree well with experiment.

There were attempts to reproduce all the basic
results of QED without second quantization. Thus,
the approach used by Barut~\cite{Barut, Barut1,Barut2}  is based on the
analysis of the self-consistent Dirac-Maxwell system
of equations studied in this paper, and, essentially, it is
purely classical. Nonetheless, the perturbation theory
and a simple regularization procedure ensure exact
reproduction of all radiative corrections usually calculated in the QED formalism.

The effect of the self-field of the electron, as noted
above, is “nonperturbative” and is not reduced to
radiative corrections. Consideration of this effect
completely alters the mathematical structure of the
model, making it essentially nonlinear. It can be
assumed that the divergences in Barut’s approach (and
correspondingly, QED) could arise due to the absence
of an exact solution to the initial classical Dirac-Maxwell system which might serve as the initial approximation in the applied perturbation theory~\cite{Pestov}.

Thus, we naturally arrive at examination of regular
solutions to the self-consistent effective nonlinear system of coupled Dirac and Maxwell equations (in the
presence of the additional external Coulomb field of
the proton). Note that such systems of equations were
considered earlier in the context of construction of an
interesting class of classical soliton models of elementary particles.

Such models are based on the well proven linear
field equations of Schrödinger, Klein–Gordon,
Dirac, and Maxwell~\footnote{Or on field structures with “natural” nonlinearity, Yang–Mills
	and/or Einstein equations.}.  In this case, the effective nonlinearity appears only as a consequence of interaction
between different fields, and the form of this interaction is well defined by the requirement of gauge invariance.

The founder of this approach is, perhaps, N. Rosen, who considered stationary spherically symmetric so-called
particle-like (i.e., everywhere regular with finite
Noether’s integrals of motion) solutions to the Klein-Gordon and Maxwell system of equations with minimal electromagnetic interaction between them in~\cite{Rosen}.

Later, this approach was extended to the (mathematically much more complicated) system of Dirac-Maxwell equations with minimal electromagnetic
interaction between fields~\cite{Wakano, Vestnik, YadPhys, DMSoliton}.  It is especially
interesting that this system is a classical analog of
operator equations of quantum electrodynamics. It
was shown, in particular, that charged fermions with
any half-integer spin can be described in a unified way
in the framework of this model~\cite{Vestnik,Thesis}. 

Note now that the real particles' properties are directly manifested if only they move in external, in particular, electromagnetic, fields. Examination of the corresponding nonstationary particle-like solutions is quite complicated.
In this regard, stationary solutions to the Dirac-Maxwell system of equations in the presence of an external Coulomb potential turn out to be physically attractive and mathematically feasible for consideration.

For hydrogen-like atoms regular solutions to a
nonrelativistic analog of the considered system, the
self-consistent system of Schrödinger–Poisson equations, were considered in~\cite{Pestov}; however, just for the
case of “pure” hydrogen ($Z=1$)  a solution had not
been found. Preliminary examination and search for particle-like (or, using modern terminology, soliton-like)
solutions in the relativistic Dirac-Maxwell model of
the hydrogen atom can be found in~\cite{Hydrogen} (see also~\cite{Leon}). In this paper, we specify more precisely and develop
the results of these previous studies.
      
\section{FIELD EQUATIONS AND BASIC PARAMETERS OF SOLUTIONS}

We consider the self-consistent system of Dirac-Maxwell equations in Minkowski space $\bf M$ with the
metric $\eta_{\mu\nu} = diag\{+1,-1,-1,-1\}$   and coordinates $\{X^\mu\},~ \mu = 0,1,2,3$. This system of equations corresponds to
the Lagrangian of the form
\be{Lagrang}
L = -\frac{1}{16\pi} F^2 + \frac{\hbar c}{2}\{i\bar \Psi \gamma^\mu \prt_\mu \Psi   - i\prt_\mu \bar \Psi \gamma^\mu \Psi - 2k\bar \Psi \Psi\} +eA_\mu \bar \Psi \gamma^\mu \Psi.
\ee
Here, $\prt_\mu:=\prt/\prt X^\mu$,~~$\{\gamma^\mu\}$  are the $4\times 4$ Dirac matrices in the canonical representation, $\Psi$  is the Dirac
bispinor, $\bar \Psi: = \Psi^+ \gamma^0$, $\{A_\mu\}$  are the 4-potentials of electromagnetic (EM) field, and $F^2 = F_{\mu\nu} F^{\mu\nu}$  is the main
invariant expressed in terms of EM field strength tensor $F_{\mu\nu} = \prt_\mu A_\nu - \prt_\nu A_\mu$. 

It is assumed that the scalar potential $A_0$ includes,
along with the regular part, the fixed singular part $A_0^{ex}=e/R$, where $e$  is the elementary charge. This part
corresponds to the external Coulomb field of the proton.

The quantities $\hbar, k$,  generally speaking, represent
some scaling coefficients whose numerical values
should be found from the comparison of the parameters of some standard solution (corresponding, for
example, to the ground state of the hydrogen atom)
with experimental values. However, it should be
expected from the agreement with the quantum theory
that these quantities correspond to the canonical constants, the Planck constant ($\hbar=h/2\pi$)  and the inverse
Compton electron length $k=mc/\hbar$, where $m$  is the
electron mass. 
  
By varying the action $S=\int L d^4 x$ over $\bar \Psi$ and $A_\mu$ and applying the Lorentz gauge condition $\prt_\mu A^\mu = 0$ to
the potentials, we obtain the field equations of
the form
\be{Eqs}
i\gamma^\mu (\prt_\mu - i (e/\hbar c ) (A_\mu+A_\mu^{ex}) \Psi - k\Psi=0, ~~\Box A^\mu = 4\pi e \bar \Psi \gamma^\mu \Psi,
\ee
where $A_\mu^{ex} = \{e/R, 0,0,0\}$. 

Now, we impose on the sought solutions of system (\ref{Eqs}) the condition of electroneutrality $\vert Q \vert = e$, i.e., the condition of equality of the absolute value of the electron
distribution charge $Q= -e \int \Psi^+\Psi d^3 X$ and the proton
charge $e$. Satisfaction of this condition is equivalent to
the following:
\be{Neutral}
\int \Psi^+\Psi d^3 X = 1,
\ee
so that it  automatically guarantees normalization of the
“wave function” of electron distribution, although
originally the field $\Psi$ can be considered classical~\footnote{In the linear problem, normalization is achieved through multiplication by an appropriate constant, which does not affect energy levels. In our nonlinear case, for the normalization condition to
	be satisfied, a nontrivial procedure is required (see Section 5 for
	details), and the parameters of normalized and nonnormalized
	solutions differ.}. 

Using scale transformations of coordinates (converting lengths to the Compton scale) and field functions of the form  
\be{scale}
X_\mu \mapsto x_\mu /k ,~ A_\mu \mapsto \frac{ke}{\alpha} a_\mu , ~\Psi \mapsto \left(\frac{k^3}{4\pi\alpha}\right)^{1/2} \psi, 
\ee
we obtain the Lagrangian 
\be{lagrang}
l = -\frac{1}{16\pi} f^2 + \frac{1}{8\pi}\{i\bar \psi \gamma^\mu \prt_\mu \psi   - i\prt_\mu \bar \psi \gamma^\mu \psi - 2\bar \psi \psi +2(a_\mu + a_\mu^{ex})\bar \psi \gamma^\mu \psi\}, ~~a_\mu^{ex}=\{\frac{\alpha}{r},0,0,0\},
\ee 
and the corresponding field equations
\be{eqs}
i\gamma^\mu (\prt_\mu - i( a_\mu+a_\mu^{ex})) \psi - \psi =0, ~~\Box a^\mu = \bar \psi \gamma^\mu \psi,
\ee
which do not contain any dimensional parameters.
The only dimensionless parameter, the fine structure
constant $\alpha:=e^2/\hbar c \approx 1/137$,  
 enters only the expression for the Coulomb potential $a_0^{ex}=\alpha/r$ and the condition of electroneutrality/normalization which takes
 the following form in terms of dimensionless variables:
\be{neutral}
\frac{1}{4\pi}\int \psi^+\psi d^3 x = \alpha.
\ee

Note that dimensional parameters corresponding
to original Lagrangian (\ref{Lagrang}) are connected with the corresponding dimensionless quantities extracted from (\ref{lagrang}) and given in Section 4 as follows:
\begin{itemize}
\item
 For the Lagrangian  
\be{lagrangC}
L=\frac{\hbar c k^4}{\alpha} l;
\ee
\item
 for the energy
\be{actionC}
W=\frac{\hbar c k}{\alpha}w \equiv \frac{mc^2}{\alpha} w;
\ee
\item 
for the total angular momentum vector
\be{angularC}
{\bf J}=\frac{\hbar}{\alpha}{\bf j};
\ee
\item
 for electric charge (if (\ref{neutral}) is satisfied)
\be{chargeC}
Q = -e;
\ee
\item
and for the magnetic moment vector 
\be{magmomC}
{\bf M} =\frac{e}{k \alpha}{\bf m}\equiv \frac{1}{\alpha}\left(\frac{e\hbar}{mc}\right) {\bf m}.
\ee
\end{itemize}

Hereinafter, our basic task is to obtain regular solutions to system of equations (\ref{eqs}) with Lorentz gauge
and electroneutrality condition (\ref{neutral}); we consider only
stationary solutions of the form
\be{station}
 \psi^T = \{\kappa^T({\bf r}),~ \chi^T({\bf r})\} e^{-i\omega t}, ~~a_\mu = \{\phi({\bf r}), \bf a({\bf r})\}, 
 \ee
 where $\kappa, \chi$  are the 2-spinors, $\phi, \bf a$ are the scalar and
 vector electromagnetic potentials, respectively, and
 the symbol `T' denotes transposition.

\section{$\alpha$-EXPANSION \\ AND THE NONRELATIVISTIC LIMIT $\alpha \rightarrow 0$}

In stationary case (\ref{station})  the basic system of equations (\ref{eqs}) takes the form
\be{eqsstat}
\begin{array}l
{\bf \Sigma}\cdot (i\nabla - {\bf a}) \chi =(\omega + \phi + \alpha/r - 1)\kappa ,\\ 
{\bf \Sigma}\cdot (i\nabla - {\bf a}) \kappa=  (\omega + \phi + \alpha/r +1)\chi , \\
\Delta \phi =\kappa^+\kappa + \chi^+\chi, \\
\Delta {\bf a} = \kappa^+{\bf \Sigma} \chi + \chi^+{\bf \Sigma} \kappa,
\end{array}
\ee
with the electroneutrality condition $(4\pi)^{-1} \int (\kappa^+\kappa + \chi^+\chi) d^3 x =\alpha$. Here ${\bf \Sigma}=\{{\bf \Sigma}_a\}, a=1,2,3$  is the set of Pauli matrices in the standard representation.

Let us now make one more transformation of coordinates and field functions~\cite{Hydrogen},  to convert length to the
Bohrian scale and expand the solutions over the small
parameter $\alpha$. Namely, let us introduce the new coordinates  $\bf X$, 2-spinors  $K,N$  and the potentials $\Phi, \bf A$ (not to be confused with the original dimensional
ones) as follows:
\be{bor}
\begin{array}l
{\bf x} = {\bf X}/(\alpha (1+\omega)), \\
\kappa = \alpha^2 (1+\omega)^{3/2} K, ~~\chi=\alpha^3 (1+\omega)^{3/2} N, \\
\phi = \alpha^2 (1+\omega)\Phi, ~~{\bf a} = \alpha^3 (1+\omega) \bf A.
\end{array}
\ee
Then system  (\ref{eqsstat}) takes the form
\be{eqsrenorm}
\begin{array}l
{\bf \Sigma}\cdot (i\nabla - \alpha^2{\bf A}) N =(- \varepsilon + \Phi +1/R) K, \\ 
{\bf \Sigma}\cdot (i\nabla - \alpha^2 {\bf A}) K=(1+\alpha^2 (\Phi + 1/R)) N  , \\
\Delta \Phi =K^+K + \alpha^2 N^+N, \\
\end{array}
\ee
while the additional equation for the magnetic potential reads
\be{eqmagn}
\Delta {\bf A} = K^+{\bf \Sigma} N + N^+{\bf \Sigma} K,
\ee
and the electroneutrality condition takes the form $(4\pi)^{-1} \int (K^+K + \alpha^2 N^+N) d^3 X =1$. Here, the “reduced
frequency” is 
\be{epsilon}
\varepsilon:= (1-\omega)/(\alpha^2 (1+\omega)), 
\ee
and it follows from the analysis of the asymptotics
(see below) that solutions  (\ref{eqsrenorm}) decreasing at infinity
can exist only for $\vert \omega \vert <1$, , so the parameter $\varepsilon$ is always
positive.

The new form of Eqs. (\ref{eqsrenorm}) is convenient because,
due to the smallness of the parameter $\alpha$, the “zero”
approximation $\alpha \rightarrow 0$ can be considered in a consistent way. Indeed, by discarding terms $\sim \alpha^2$,  we obtain for the first three Eqs. (\ref{eqsrenorm})  and the electroneutrality
condition
\be{principl}
\begin{array}l
i{\bf \Sigma}\cdot \nabla K =N ,\\ 
i{\bf \Sigma}\cdot \nabla N =  (- \varepsilon + \Phi +1/R) K , \\
\Delta \Phi =K^+K , \\
(4\pi)^{-1}\int K^+K d^3 X =1,
\end{array}
\ee
System (\ref{principl}) is closed, it contains the only parameter $\varepsilon$ to be defined, and it does not include terms with
the magnetic potential. The latter can be defined from
the prefound solutions to limiting system  (\ref{principl}) by integration of Eq (\ref{eqmagn}). In this case, system (\ref{principl}) after substituting the expression for the 2-spinor $N$ from the
first equation to the second equation is reduced to one
second order equation for the main 2-spinor $K$. After
simple transformations, we obtain the following final
form of the system in the leading approximation with
respect to $\alpha$:
 \be{eqsfin}
  \begin{array}l
 \Delta K =  (\varepsilon - \Phi -1/R ) K, \\
\Delta \Phi =K^+K , \\
(4\pi)^{-1}\int K^+K d^3 X =1. 
\end{array}
 \ee
The structure of system  (\ref{eqsfin})  is close (but not identical) to the self-consistent system of Schrödinger–
Newton equations which attracted attention earlier
thanks to the ideas of Diosi~\cite{Diosi}, Penrose~\cite{Penrose} and others, on the possible self-gravitating nature of reduction
of the particle wave function mentioned above. In particular, the spectrum of stationary spherically symmetric regular solutions to the Schrödinger–Newton
system was numerically obtained in~\cite{Bernstein, Tod}.

The Schrödinger-Newton system differs from system  (\ref{eqsfin})  by the absence of an external Coulomb
(Newtonian) potential and the sign of potential in the
first equation of type (\ref{eqsfin})  (which is connected with
the different signs of the gravitational and electrostatic
interactions). It can be easily seen that in reality (\ref{eqsfin}) represents a self-consistent system of Schrödinger-Poisson equations (written for the case of stationary
solutions), so that the transition to the limit $\alpha \mapsto 0$ and system (\ref{principl}) or (\ref{eqsfin})  represent, in essence,  the nonrelativistic approximation of the original Dirac-Maxwell system.

Regular solutions the Schrödinger-Poisson equations are in fact the main
subject of the study below. As for the problem of finding
relativistic corrections, it is quite complicated, since, as
will be shown later, in the general case it results in an
infinite chain of equations for radial functions.

It should be noted that systems of equations similar
to  (\ref{eqsfin})  were proposed more than once as a physically
justified alternative to the linear Schrödinger equation.
Schrödinger himself discussed the need to consider
the “spatial distribution” of the electron charge,
including the case of multielectron atoms~\cite[p. 116]{Schrod}, and the
possibility of introducing a “neutralization potential” for this purpose~\cite[p. 175]{Schrod}, i.e., the potential of the self-field
of electrons. Detailed examination of these issues can
be found in~\cite{Vlasov}.

To the best of our knowledge, the first attempts to
obtain the regular solutions to system  (\ref{eqsfin})  and the corresponding binding energies were made in ~\cite{Hydrogen}  and ~\cite{Pestov}. In ~\cite{Pestov} these solutions were found for lowest s-states
(with  $n=1,2,3$) for different nuclear charge  $Ze$ of
hydrogen-like atoms; the iteration method was used,
and wave functions of the canonical linear quantum
mechanical problem were taken as the first approximation.

The obtained energy level shifts, as expected,
turned out large, and in order to agree with the
observed values, it was proposed to use an analog of
renormalization. However, the correctness of the procedure of finding solutions raises doubts~\footnote{It is not clear whether the wave function normalization was preserved during iterations. Besides, the expression for the binding
	energy, along with the term corresponding to the eigenvalues of
	the “customary” quantum mechanical energy $\varepsilon$,  should include
	the self-energy of the electron field, see (\ref{energyB}) below.}. Anyway, the
shift was not determined for the most important case
of hydrogen itself ($Z=1$).

A more reliable method of the numerical solution
for system~(\ref{eqsfin}), together with the variational method
(in the class of the same trial functions as in the linear
quantum mechanical problem), was used in (\cite{Hydrogen, Leon}). Moreover, the solutions found therein can be used to
find the magnetic potential and the electron angular
momentum in different states of the hydrogen atom by
integrating Eq. (\ref{eqmagn}).

\section{AXIALLY SYMMETRIC SOLUTIONS: SEPARATION OF ANGULAR DEPENDENCE AND THE PROBLEM OF ENTANGLEMENT OF  HARMONICS}

Strictly speaking, the system of equations (\ref{principl}) under study has no spherically symmetric solutions
due to the spinor nature of the functions  $K,N$ and the
presence of magnetic potential of the electron field
calculated from additional Eq. (\ref{eqmagn}). System (\ref{eqsfin}) itself,
however, can be reduced to an equation for one spherical harmonic corresponding to the main 2-spinor $K$ and the corresponding potential $\Phi$. Indeed, by assuming the potential spherically symmetric, it can be seen
that the angular dependence in the equations for the
spinor functions in  (\ref{principl})  can be described by any pair of
corresponding spherical spinors, i.e., similarly to the
linear problem of quantum mechanics.
On the other hand, the corresponding electric
charge density $\sim K^+ K$ in the right-hand part of (\ref{principl}) is
not necessarily spherically symmetric, so the system
turns out to be unclosed and requires additional harmonics of the potential (corresponding to the quadrupole and higher even moments). The inevitable appearance of these harmonics, in turn, requires
introduction of the corresponding pairs of higher
order spherical spinors, and so on. This situation
results in the requirement to consider an infinite chain
of equations for radial functions, and it is difficult to
substantiate the cut in this chain because of the
absence of any small parameter. This problem was
considered in detail in~\cite{Thesis}. 

The only possibility to obtain a closed system of
equations is the choice of such initial pairs of spherical
spinors for which the corresponding electric charge
density is spherically symmetric (below, we denote the
radial variable $R \mapsto r$ for simplicity) 

Thus, the choice of the angular dependence of the
field functions is limited by the following two possibilities corresponding, from the point of view of canonical quantum mechanics, to s- and p-states with the
eigenvalues of the operator of orbital angular momentum $l=0,1$ respectively:

ansatz {\bf А}: 
\be{sysA}
\Phi = \phi(r),~~ K=k(r) \left( \begin{array}l
1 \\ 
0
\end{array} \right), ~~ N=i n(r) \left( \begin{array}l
\cos {\theta} \\ 
\sin {\theta} \ e^{i\varphi}
\end{array} \right),
\ee
while, after the replacement $K \leftrightarrow N$, we have ansatz {\bf B}:
\be{sysB}
\Phi = \phi(r),~~ K=i k(r) \left( \begin{array}l
\cos {\theta} \\ 
\sin {\theta} \ e^{i\varphi}
\end{array} \right), ~~
N=n(r) \left( \begin{array}l
1 \\ 
0
\end{array} \right). 
\ee

Note that two more substitutions are possible,
along with the two angular dependences given above: 
\be{sysA2}
\Phi = \phi(r),~~ K=k(r) \left( \begin{array}l
0 \\ 
1
\end{array} \right), ~~ N=i n(r) \left( \begin{array}l
- \sin {\theta} \ e^{- i\varphi} \\ 
~~~\cos {\theta} 
\end{array} \right),
 \ee 
 or  
\be{sysB2}
\Phi = \phi(r),~~ K=i k(r) \left( \begin{array}l
- \sin {\theta} \ e^{- i\varphi}\\ 
~~~\cos{\theta}
\end{array} \right), ~~
N=n(r) \left( \begin{array}l
0 \\ 
1
\end{array} \right), 
\ee
conjugate to the ansatzes $\bf A$ and $\bf B$~ respectively. It can
be easily verified that they result in the same form of
system (\ref{eqsfin}), and the corresponding solutions differ
only by the signs of projections of the magnetic
moment and the total angular momentum. Therefore,
we will not consider them separately.
For both substitutions $\bf A$ and $\bf B$ main system of
equations (\ref{principl}) and system (\ref{eqsfin}) following from it turn
out consistent and closed. Namely, for $\bf A$ we obtain:
\be{sysAA}
\begin{array}l
k^\p = n, ~~n^\p + (2/r) n = (\varepsilon - \phi - (1/r)) k \\
\mapsto k^{\p\p} + (2/r) k^\p = (\varepsilon - \phi - (1/r)) k, \\
\phi^{\p\p} + (2/r) \phi^\p = k^2, \\
\int k^2 r^2 dr =1,\\
\lambda^{\p\p} + (2/r) \lambda^\p - (2/r^2) \lambda =  2 k  k^\p,  
\end{array}
\ee
 and for  $\bf B$ we obtain:
 \be{sysBB}
\begin{array}l
k^\p  + (2/r) k= - n, ~~n^\p = - (\varepsilon - \phi - (1/r)) k \\
\mapsto k^{\p\p} + (2/r) k^\p - (2/r^2) k = (\varepsilon - \phi - (1/r)) k, \\
\phi^{\p\p} + (2/r) \phi^\p = k^2, \\
\int k^2 r^2 dr =1,\\ 
\lambda^{\p\p} + (2/r) \lambda^\p - (2/r^2) \lambda = - 2 k (k^\p +(2/r) k), 
\end{array}
\ee
where $(\p)$ denotes differentiation over the radial variable $r$, and integration hereinafter is performed over
the semi-infinite interval $[0, \infty)$. 

The additional last equation in (\ref{sysAA}) and (\ref{sysBB}) represents the equation for the $\varphi$--component of the magnetic potential
$A_\varphi = \lambda (r) \sin{\theta}$, corresponding to  (\ref{eqmagn}). There, the $r, {\theta}$-components of current densities corresponding to substitutions (\ref{sysA}), (\ref{sysB}), are identically equal
to zero. Therefore, we can set for the corresponding
magnetic potential components $A_r = A_\theta = 0$. 

The expression for the magnetic moment of the
sought regular solutions to system (\ref{sysAA}) can be easily
obtained. Indeed, by integrating the last equation with
the weight $r^3$ and considering the electroneutrality
condition and the asymptotic behavior $\lambda \sim \mu/r^2$, where $\mu$ is the magnetic moment of the distribution,
we obtain
\be{magmom}
\int ((\lambda^\p - \lambda /r) r^3)^\p dr = -3\mu = \int (k^2)^\p r^3 dr = -3 \int k^2 r^2 dr = -3,
\ee
which yield $\mu=1$,  or in terms of dimensional units (\ref{magmomC}), with consideration of transformation (\ref{bor}), 
\be{magmomA}
\vert {\bf M}_{\bf A}\vert = \left( \frac{e\hbar}{mc}\right) \frac{1}{1+\omega} \approx \frac{e\hbar}{2mc}.
\ee  
Indeed, since, due to (\ref{epsilon}), 
\be{unomega}
1+\omega=2/(1+\varepsilon \alpha^2)\approx 2 (1-\varepsilon \alpha^2),
\ee
the magnetic moment, to extremely small corrections $\sim \alpha^2$,  has the canonical Dirac value~\footnote{It will be seen below that the numerical value of the parameter $\varepsilon$ for all regular solutions is very small ($\varepsilon << 1$), so that the deviation of the magnetic moment from the Dirac value is smaller by
	many orders of magnitude than the observed and QED-predicted value of $(\alpha/2\pi)$}. 

Performing the same operations of integration and
dimensionality restoration for the magnetic moment
of solutions of the second class (\ref{sysBB}), we obtain $\mu = 1/3$,  so instead of (\ref{magmomA}) we now have 
\be{magmomB}
\vert {\bf M}_{\bf B} \vert  \approx  \frac{1}{3}\left(\frac{e\hbar}{2mc}\right). 
\ee  
This magnetic moment value seems quite unexpected
and interesting. However, it is difficult to propose a reasonable physical interpretation for such a value.

Now let us find the representation for (dimensionless) energy and angular momentum of regular solutions. Using the known expression for the symmetric
energy-momentum tensor $T^{(\mu\nu)}$,  corresponding to the
system of equations of type (\ref{eqs}),
\be{tensorEM}
\begin{array}l
T^{(\mu\nu)}=(4\pi)^{-1}(- \eta_{\beta\rho} F^{\mu\beta}F^{\nu\rho}+ \frac{1}{4} 
\eta^{\mu\nu} F^2) + \\ (16\pi)^{-1} ( i\bar \psi \gamma^\mu \prt^\nu \psi  - i\prt^\nu \bar \psi \gamma^\mu \psi + i\bar \psi \gamma^\nu \prt^\mu \psi  - i\prt^\mu \bar \psi \gamma^\nu \psi +2a^\mu \bar \psi \gamma^\nu \psi+ 2a^\nu \bar \psi \gamma^\mu
 \psi)
 \end{array}
\ee
we obtain for the energy of regular solutions
\be{energyw}
w =\frac{\omega}{4\pi} \int \psi^+\psi d^3 x - \frac{1}{8\pi} \int (\nabla \phi)^2 d^3 x
\ee
On the other hand, considering the so-called
“Laue’s theorem”~\cite{Laue} valid for stationary regular
solutions 
 ($\int T^{(aa)} d^3 x =0,~~a=1,2,3$)  we have
$w = \int T^\mu_\mu d^3 x$,  and for energy (\ref{energyw}) we obtain~\cite{Vestnik,Thesis} another representation in terms of the (easily calculated if the field equations are used) trace of the
energy-momentum tensor $T^\mu_\mu$,
\be{energyw2}
w = \frac{1}{4\pi}\int \bar \psi \psi d^3 x .
\ee
Going over to the Bohrian scale (\ref{bor}), considering
the electroneutrality condition and expression (\ref{unomega}) for $\omega$ and  restoring then the dimensionality according
to (\ref{actionC}), we obtain instead of (\ref{energyw})
\be{energyBor}
W = mc^2 \left(\frac{1-\varepsilon \alpha^2}{1+\varepsilon \alpha^2} - \frac{\alpha^2}{1+\varepsilon \alpha^2} \int (\phi^\p)^2 r^2 dr \right) = mc^2 - W_B .
\ee
Here, the binding energy $W_B$ up to corrections of
order $\alpha^2$ is
\be{energyB}
W_B = 4 R \left(\varepsilon + \frac{1}{2} \int (\phi^\p)^2 r^2 dr)\right),  
\ee  
where $R:=mc^2 \alpha^2/ 2 \equiv me^4/2\hbar^2$ corresponds (for $m$ being equal to the electron mass) to the Rydberg constant.
The reduced frequency $\varepsilon$ and the energy of the electric
self-field of the electron (second term in (\ref{energyB}))  should
be determined from the solution of the boundary value
problem for the system of equations (\ref{sysAA}) or (\ref{sysBB}).

Similarly, another representation for energy (\ref{energyw2}) after the transition to dimensional quantities yields the
following expression for the binding energy $\tilde W_B$
\be{energy2B}
\tilde W_B = 4R \int n^2 r^2 dr
\ee
The equality of expressions  (\ref{energyB}) and (\ref{energy2B}) represents an integral identity satisfied for any regular
solutions to system (\ref{sysAA}) or (\ref{sysBB}); it was used for estimation
of the accuracy of the obtained numerical solutions of the
corresponding boundary value problem (see Section 5).

Finally, let us derive the representation of the total
angular momentum of regular solutions. Due to the symmetric character of energy-momentum tensor (\ref{tensorEM}) the
tensor of total angular momentum $M^{\mu[\nu\lambda]}$ has the form 
\be{tensang}
M^{\mu[\nu\lambda]}:=x^\nu T^{(\mu\lambda)} - x^\lambda T^{(\mu\nu)}, 
\ee
and the conserved 3-vector of angular momentum $\bf j$ is
given by the expression:
\be{spin}
{\bf j} = \int {\bf r}\times {\bf p}~ d^3 x,   
\ee                                                            
in which the density of the momentum vector ${\bf p}^c$ is
determined by the components  $T^{(0c)}$ of the energy-momentum tensor
\be{momentum}
{\bf p}:= \frac{1}{4\pi} ({\bf E} \times {\bf H}) - \frac{1}{8\pi} ({\bf a}\psi^+\psi +(\omega+\phi)\psi^+{\bf s} \psi +\frac{i}{2}(\psi^+\nabla \psi -\nabla \psi^+ \psi)) , 
\ee
where ${\bf s}^c :=i\gamma^0 \gamma^c $. 
Using field equations  (\ref{eqs}) and integrating by parts, we obtain
\be{ja}
{\bf j} = \frac{1}{8\pi}\int {\bf r}\times (i\nabla \psi^+\psi-i\psi^+\nabla \psi)~ d^3 x + \frac{1}{8\pi}\int \psi^+ {\bf \Lambda} \psi ~d^3 x,
\ee
where $ {\bf \Lambda}^c :=  (i/2) \varepsilon_{abc} \gamma^a \gamma^b$.  In the case of axial symmetry, the only nonzero component of the angular
momentum takes the form
\be{spinax}
j={\bf j}_3 = \frac{1}{4\pi}\int \psi^+ (-i\frac{\prt}{\prt \varphi} +\frac{1}{2}\Lambda_3) \psi~d^3 x. 
\ee
After transition to the nonrelativistic limit we
obtain the following ezpression for solutions of both
classes {\bf A} and {\bf B}:
\be{spinnonrel}
j = \frac {1}{2} \int k^2 r^2 dr =\frac{1}{2}, 
 \ee 
 by considering that the electric charge is equal to the
 elementary charge. In terms of dimensional units, as
 expected~\footnote{The relation between the total angular momentum and electric
 	charge of regular axially symmetric solutions has a universal
 	character, and in the case of absence of an external potential, it
 	was obtained in ~\cite{Thesis} in the form  $J = (Q/e) m\hbar/2  , ~~m=1,2,...$. Thus, the models of the considered class can, in principle,
 	describe charged fermions with any half-integer spin.}, we have~$J=\hbar/2$. For solutions determined
 by ansatzes (\ref{sysA2}), (\ref{sysB2}), the projection of the angular
 momentum $J$ has the opposite sign, $J=- \hbar/2$.  
   
\section{ NUMERICAL STUDIES AND BOHRIAN BINDING ENERGY SPECTRUM}
Below, we study everywhere regular solutions to
system (\ref{sysAA})  which, after introduction of the “reduced”
potential
\be{reducepot}
 U:=\varepsilon - \phi,
 \ee
and ignoring the additional equation for the magnetic
potential $\lambda$ takes the form
\be{sysAfin}
\begin{array}l
k^\p = n, ~~n^\p + (2/r) n = (U - (1/r)) k \\
\mapsto k^{\p\p} + (2/r) k^\p = (U - (1/r)) k, \\
U^{\p\p} + (2/r) U^\p = - k^2 .
\end{array}
\ee
The asymptotics of the solutions of interest have
the form
\be{aszero}
k\sim \beta (1-r/2+...),~~ n\sim -\beta /2 +...,~~ U\sim \delta - \beta^2 r^2 /6+..., 
\ee
for small $r\sim 0$, and
\be{asinf}
k\sim A e^{-\sqrt{\varepsilon} r},~~ n \sim - \sqrt{\varepsilon} k, ~~ U\sim \varepsilon + q/r, ... 
\ee 
for large  $r \rightarrow \infty$. 

Assuming the existence of such solutions with an
arbitrary number of nodes $N=0,1,2,...$ with respect to
the main function $k(r)$, we arrive at the nonlinear
boundary value problem for the “eigen” values: for
given $\delta = U(0)$ it is necessary to fit such $\beta = k(0)$, for
which the function  $k(r)$ decreases, asymptotically
approaching zero, at large distances from the center. 
In this case, the potential  $U(r)$ slowly
decreases by itself, according to asymptotics (\ref{asinf}), which is
used to determine the eigen value  $\varepsilon$. 

We use the known (see, e.g.,~\cite{Tod}) “bracketing”
method, reducing the interval of $\beta$ values according to
the following criterion for numerical integration. If the
function $k(r)$ “blows up” and begins to increase, $n(r_0)=k^\p(r_0) >0$,  for some value of the argument, $r=r_0$, the initial value of $\beta$ should be reduced.  If, conversely, (for the nodeless solution with $N=0$) for
some $r_0$ the function $k(r)$ becomes negative,  $k(r_0)<0$, the value of $\beta$ should be increased. As a result, we
determine the approximate eigen value $\beta_0$, for which
the blowing up of the solution one way or another
takes place only at a sufficiently large $r_0$. By finding the value
of $q\approx -r_0^2 U^\p(r_0)$, at this point, we determine the normalization integral, the charge $q\approx \int k^2 r^2 dr$.

After that, the procedure is repeated for another
initial value $\delta$ of the potential $U$,  resulting in the regular solution with another normalization integral $q$. Finally, we obtain the approximate values of  $\beta_0, \delta_0$, corresponding to the normalized (and therefore, electrically neutral) nodeless regular solution to system (\ref{sysAfin}), with the form shown in Fig. 1 (cut-off radius $r_0\sim 90)$).The form of the corresponding electrostatic potential
 $U$ is shown in Fig. 2.
	\selectlanguage{English}
\begin{figure}[h!]
	\center{\includegraphics[scale = 0.5]{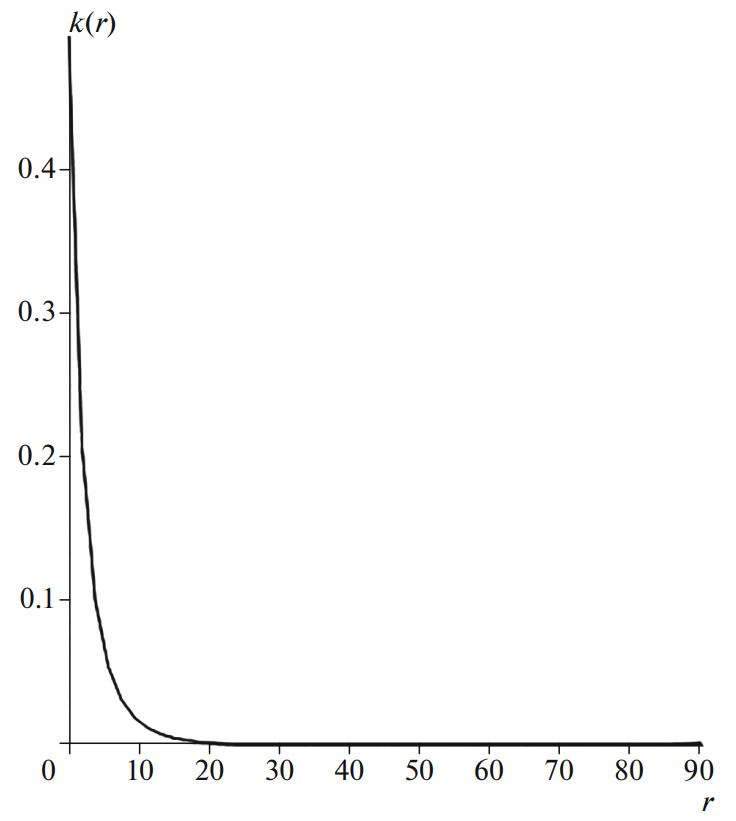}}\\  
	\large\caption{ Form of the nodeless regular solution corresponding to the ansatz \textbf{A}}         
	\label{fig:1}
\end{figure}

\begin{figure}[h!]
	\center{\includegraphics[scale = 0.5]{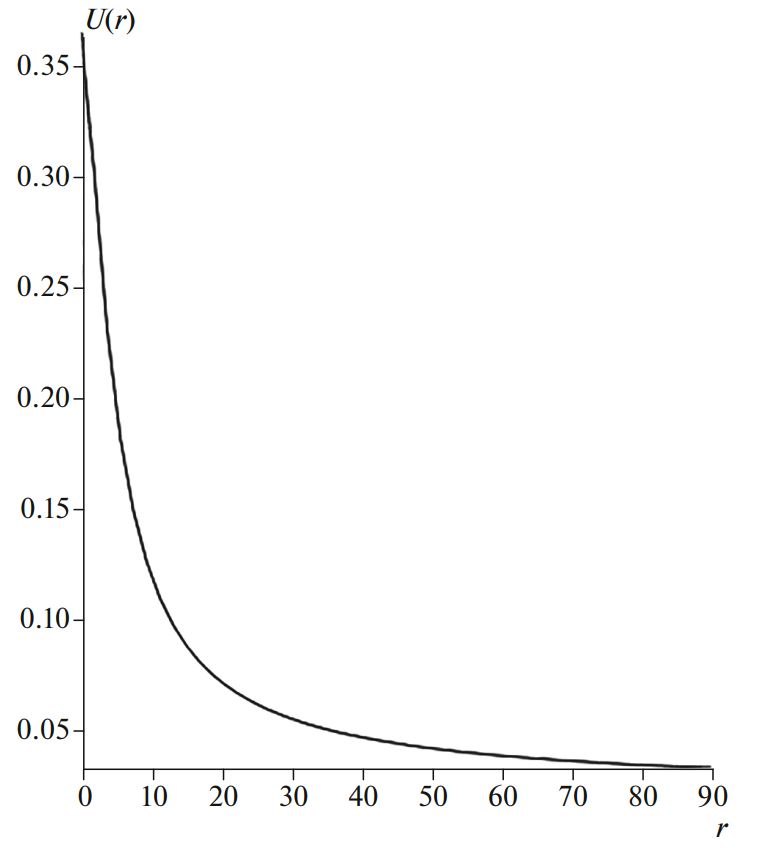}}\\  
	\large\caption{ Form of reduced electric potential for the nodeless
		solution corresponding to the ansatz  \textbf{A}}         
	\label{fig:2}
\end{figure}
	\selectlanguage{Russian}
Then, for the found $\beta_0$ we integrate the equation for
magnetic potential in (\ref{sysAA}),  choosing such value of $\gamma_0$ for the ``derivative $\lambda^\p$ at zero'', for which the growing
term in the asymptotic $\lambda\sim Hr + \mu/r^2$ approximately
vanishes at large distances from the center; then we
determine the magnetic moment of the distribution $\mu=r_0^2 \lambda(r_0)$, making sure that with good accuracy $\mu\approx 1$  (see (\ref{magmomA})).  The sought form of the magnetic
potential for the nodeless solution is shown in Fig. 3.

\selectlanguage{English}
    \begin{figure}[h!]
	\center{\includegraphics[scale = 0.60]{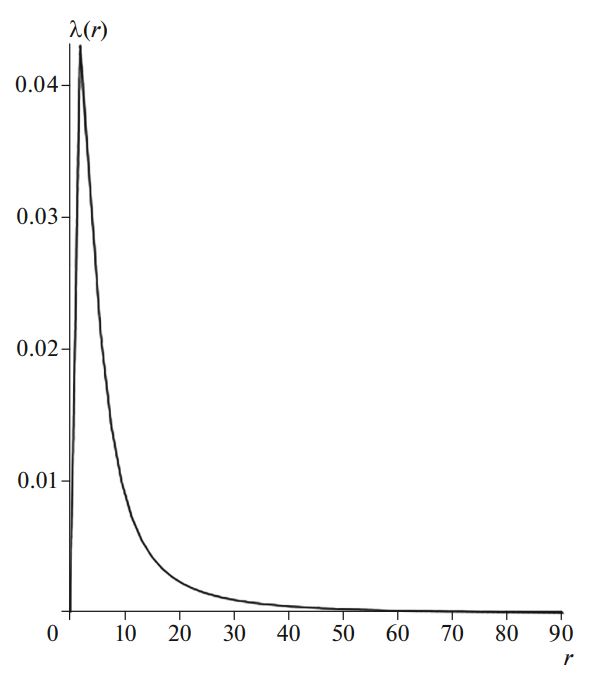}}\\  
	\large\caption{Form of a magnetic potential for the nodeless solution for the ansatz  \textbf{A} }         
	\label{fig:3}
\end{figure}
\selectlanguage{Russian}

Eventually, using all found values of $\{\beta_0, \delta_0, \gamma_0\}$,  we
find the integral characteristics of the nodeless solution by
numerical integration, and verify the accuracy of their
determination by satisfaction of the integral identities
(compare, for example, (\ref{energyB}) and (\ref{energy2B})). Finally, using the
potential  $U$ asymptotic, we determine the eigen value for
the “reduced frequency” $\varepsilon \approx U(r_0) +r_0 U^\p(r_0)$, and then
calculate the binding energy using formula ({\ref{energyB}).

   It can be easily seen that a similar procedure can be
   applied for finding regular solutions with an arbitrary
   number of given nodes $N$ of the main function $k(r)$. To this end, it is necessary to narrow down the corresponding interval of $\beta_N$ values, allowing the function $k(r)$ to change sign $N$ times. Figure 4 shows the normalized regular solution with the number of nodes $N=2$, as an example.

Table 1 gives the numerically calculated parameters
and characteristics of normalized regular solutions to
system (\ref{sysAfin})  (and corresponding solutions to (\ref{sysAA})) with
the number of nodes $N=0\div 10$. It can be seen from 
 Table 1 and Fig. 5 that the values of the binding energy $W_B$  reproduce the canonical Bohrian distribution with
 good accuracy,
\be{numbohr}
W_n= W /n^2, ~~n=N+1=1,2, ... .
\ee

The next column of Table 1 shows the corresponding values of the binding energy $W^{var}$, obtained by the
variational method. Variation was performed in the
class of trial functions corresponding to the canonical
linear quantum mechanics problem (when the potential of the self-field  $\phi(r)$ is not included in the first of
Eqs. (\ref{sysAfin}) and $U(r)=\varepsilon=const$):
\be{Lagerr}
k_N =const \cdot L_N^{(1)}(x) e^{-x/2},~~x:=r/N, ~~N=0,1,2,...,
\ee
where $L_N^{(1)}(x)$ -- are the generalized Laguerre polynomials of power  $N$ as a function of $x$, and the eigen values of the corresponding linear problem are
\be{eigenlin}
\varepsilon_N=\frac{1}{4(N+1)^2}. 
\ee
 The corresponding trial functions $\phi_n(r)$ for the
 eigen field potential were found by integration of the
 second equation in (\ref{sysAfin}) with the right-hand part corresponding to (\ref{Lagerr}), so that
\be{potLagerr}
\phi_N^{\p\p} + \frac{2}{r}\phi_N^\p = k_N^2 .
\ee
The corresponding action functional was varied
over two parameters of trial functions (\ref{Lagerr}) and (\ref{potLagerr}), namely, the amplitude and scale parameters. Good
agreement of the results obtained using the numerical
and variational methods can be explained by the fact that
inclusion of the electron self-field potential does not
qualitatively change the behavior of the functions and their asymptotics both near zero and for large values of
radial coordinate.   

For solutions with a large number of nodes $N=50~(n=51)$ and $N=100~ (n=101)$ the calculations were performed using the variational method
alone. In this case, the binding energy still corresponds to Bohr’s law (\ref{numbohr}) with good accuracy.. 

It can be seen from Table 1 (see the last column)
that the average value of the constant $W$ in (\ref{numbohr})  is
approximately $W\approx 0.120$. The spectrum of dimensional binding energy for  (\ref{energyB}) or, equivalently, (\ref{energy2B}) takes the form
\be{dimbohr}
W_B = 4R W/n^2 \approx 0.48 R/n^2.
\ee
Therefore, the effective Rydberg constant $R$ is
approximately twice as small as the experimentally
observed one.
The spectrum of regular solutions with angular
dependence corresponding to the ansatz $\bf B$, (see (\ref{sysB})) is
found similarly by numerical and variational methods.
In this case, system for the radial functions (\ref{sysAfin}) is
replaced by the following system if Eq.  (\ref{principl})  is used:
\be{sysBfin}
\begin{array}l
k^\p +(2/r) k = - n, ~~n^\p  = - (U - (1/r)) k \\
\mapsto k^{\p\p} + (2/r) k^\p - (2/r^2) k = (U - (1/r)) k, \\
U^{\p\p} + (2/r) U^\p = - k^2, \\
\end{array}
\ee
with the asymptotics of the form 
\be{asymB}
k\sim \beta r - ..., ~~n \sim -3 \beta + ..., ~~ U \sim \delta -\frac{1}{20} \beta r^4    
\ee
for $r \sim 0 $;  for $r \rightarrow \infty$ we have
\be{asymB2}
k \sim A e^{-\sqrt{\varepsilon} r }, ~~ n \sim\sqrt{\varepsilon} k, ~~ U \sim \varepsilon + q/r .  
 \ee   
 \selectlanguage{English}
 \begin{figure}[h!]
 	\center{\includegraphics[scale = 0.55]{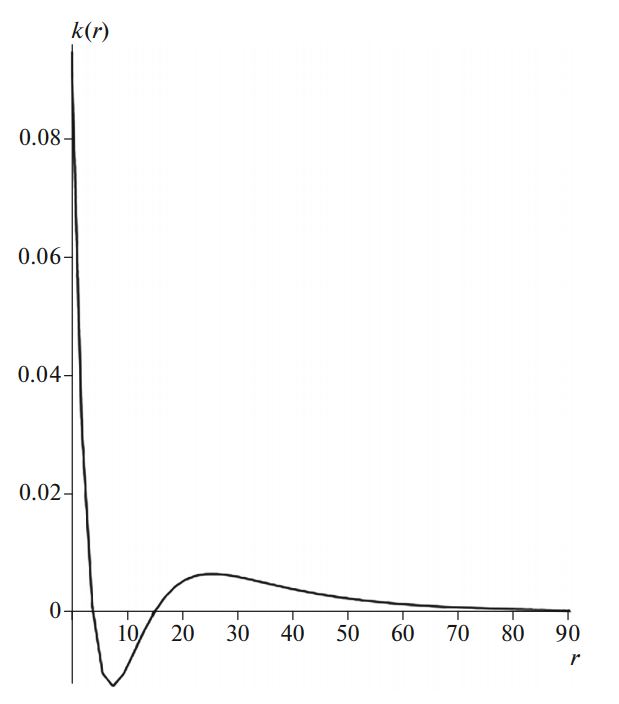}}\\  
 	\large\caption{	Form of two-node (N = 2) regular solution corresponding to the ansatz \textbf{A} }         
 	\label{fig:4}
 \end{figure}
 \selectlanguage{Russian}
Variation was performed in the class of trial functions $k(r)$, corresponding to the particular value of the
“orbital” quantum number  $l=1$ of solutions of the
linear canonical quantum mechanical problem,
namely, ($2l+1=3$):
\be{canonB}
k_N =const \cdot L_N^{(3)}(x) e^{-x/2},~~x:=r/N, ~~N=0,1,2,..., 
\ee
for which the eigen values are
\be{eigenB}
\varepsilon_N=\frac{1}{4n^2}, ~~~n = N+l+1 = N+2 ,  
\ee
while the eigen potential was determined by trial functions found by integrating former Eq.  (\ref{potLagerr}) with the
right-hand part defined now by (\ref{canonB}).

Table 2 gives the results of the numerical and variational studies of the spectrum of ground and excited
states of regular solutions to system (\ref{sysBfin}), which again
mutually prove each other.

	\selectlanguage{English}
\begin{figure}[h!]
	\center{\includegraphics[scale = 0.45]{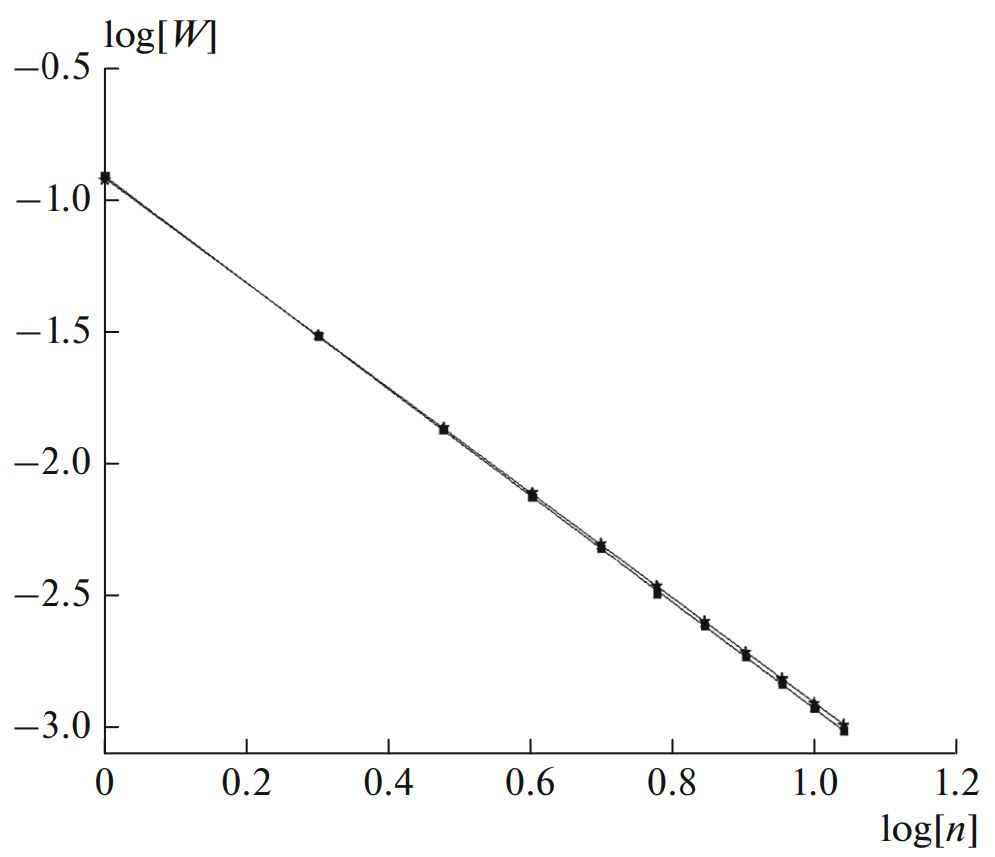}}\\  
	\large\caption{Binding energy $W_n$ of regular solutions for the
		ansatz {\bf A}   as a function of the “principal quantum number” $n=N+1$  according to (squares) the numerical and
		(stars) variational methods}           
	\label{fig:5}
\end{figure}

\begin{figure}[h!]
	\center{\includegraphics[scale = 0.45]{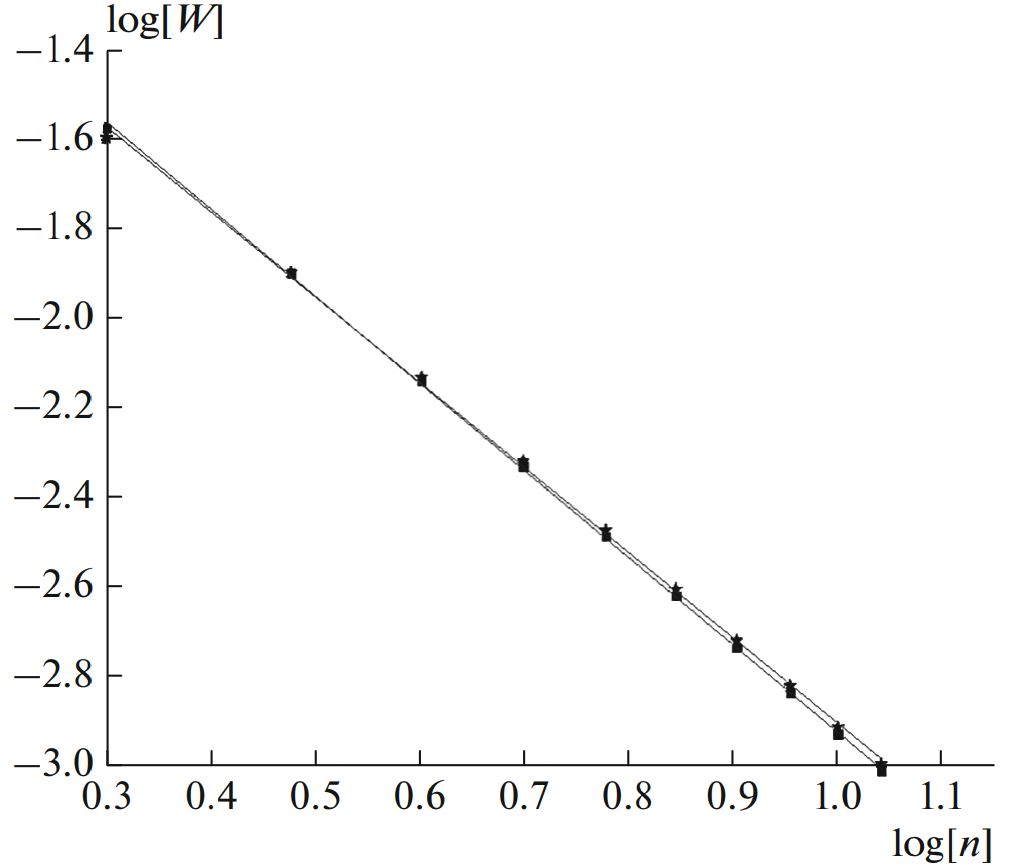}}\\  
	\large\caption{Binding energy $W_n$ of regular solutions for the
		ansatz {\bf B}   as a function of the “principal quantum number” $n=N+2$  according to (squares) the numerical and
		(stars) variational methods}         
	\label{fig:6}
\end{figure}
\selectlanguage{Russian}

It follows from this table
and Fig. 6 that the Bohrian distribution
\be{numbohrB}\
W_n = W/n^2, ~~~n=N+2=2,3,...
\ee
is again satisfied with a good accuracy, especially for
solutions with a large number of nodes $N$. Moreover,
the quantity $W$ again has the value $W\approx 0.120$, so the
effective Rydberg constant turns out universal,
approximately equal to its value for the ansatz $\bf A$ (nearly two times smaller than its experimental value).

\selectlanguage{English}
\begin{figure}[h!]
	\center{\includegraphics[scale = 0.55]{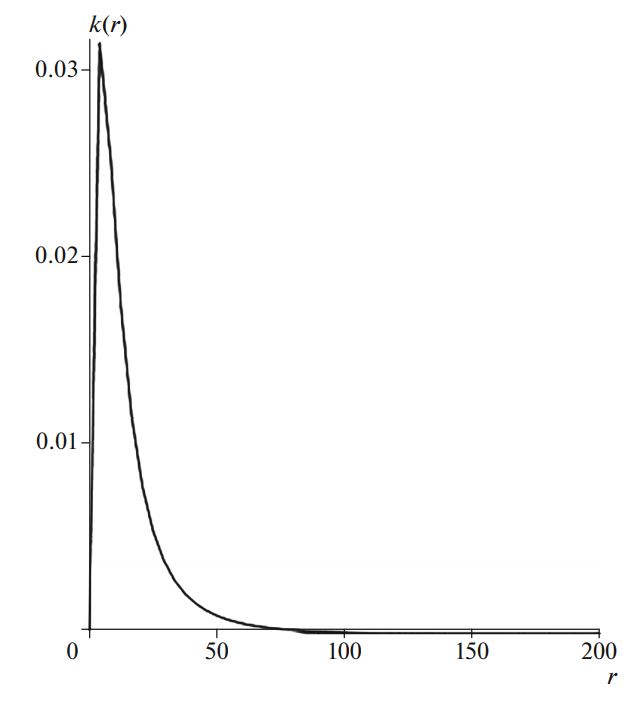}}\\ 
	\large\caption{Form of the nodeless regular solution for the ansatz  {\bf B}}         
	\label{fig:7}
\end{figure}

\begin{figure}[h!]
	\center{\includegraphics[scale = 0.55]{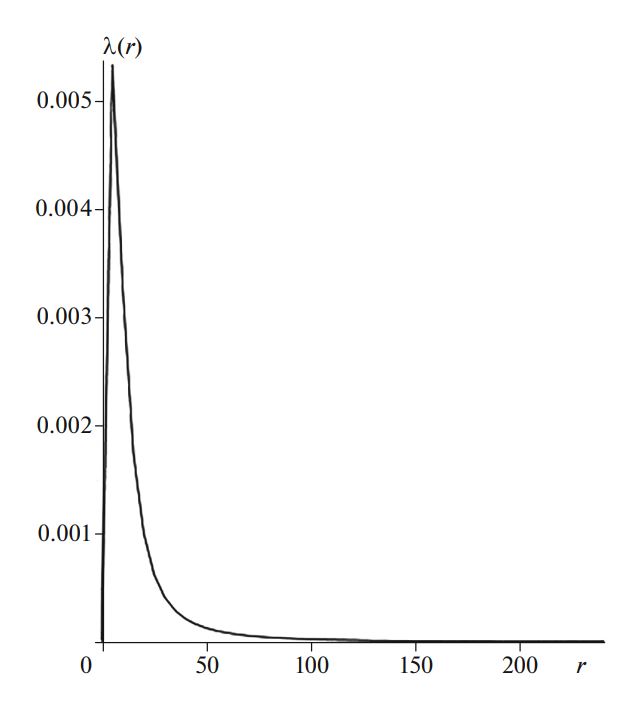}}\\  
	\large\caption{Form of a magnetic potential for the nodeless solution corresponding to the ansatz {\bf B}}         
	\label{fig:8}
\end{figure}
\selectlanguage{Russian}

The form of nodeless solution to system of equations  (\ref{sysBfin}) is given in Fig. 7 (the cut-off radius is $r_0\sim 200 $). The corresponding magnetic potential is
shown in Fig. 8.
Note finally that the values of energies of the states of
classes  {\bf A} and {\bf B} with the same value of the “principal quantum number” $n=N+l+1$  are close to each other,
especially for large $n$. Nonetheless, the difference between
them, in particular, between the energies of analogs of
states $2s_{1/2}$ and $2p_{1/2}$ (for $n=2$,~$l=0,1$ respectively)  exceeds the experimentally observed one (corresponding to the “Lamb shift”) by many orders of magnitude. \\

  Table 1. Parameters and characteristics of regular solutions for ansatz {\bf A}. 
  $$                     
  \begin{array}{lccccccc}
  n & \beta=k(0) & \delta=U(0) & \varepsilon=U(\infty) & W_n  &  W_n^{var} & W_n\cdot n^2 & W_n^{var}\cdot n^2\\ [7pt]
  1 & 0.492787  & 0.365947 & 0.02311  & 0.1218 & 0.1182 & 0.1218 & 0.1182 \\
  2 & 0.17379 &  0.09042 & 0.004515 & 0.03030 & 0.03056 & 0.1212 & 0.1222 \\
  3 & 0.09471 & 0.04005   & 0.001338  &  0.01335 & 0.01366 & 0.1201 & 0.1229\\ 
  4 & 0.06135 & 0.02243   & 0.0006103    & 0.007443  & 0.007696 & 0.1191 & 0.1231\\
  5 & 0.04382 & 0.01431 &  0.0003339  & 0.004757 & 0.004930 &  0.1189 & 0.1233\\
  6 &  0.03333 & 0.009921  &  0.0002028 &  0.003200 & 0.003425 &  0.1187 & 0.1233\\
  7 & 0.02639 & 0.007278   & 0.0001344 &  0.002418 & 0.002517 & 0.1185 & 0.1233\\ 
  8 & 0.02158 & 0.005366  &  0.00009089 & 0.001849 & 0.001927 & 0.1183 & 0.1234\\
  9 & 0.01808 & 0.004394   & 0.00006706 & 0.001461 & 0.001523 & 0.1183 & 0.1234\\ 
  10 & 0.01542 & 0.003556 & 0.00004866 & 0.001182 & 0.001234 & 0.1182  & 0.1234 \\ 
  11 & 0.01336 & 0.002937  & 0.00003744 & 0.0009755 & 0.001020 &  0.1180 & 0.1234\\
  ... & ... & ... & ... & ... &  ... & ....& .... \\[7pt]
  51 & ..... & .......& ...... & ...... & 4.745\cdot 10^{-5} & .... & 0.1234 \\
  101 & ..... & .......& ...... & ...... & 1.210\cdot 10^{-5} & .... & 0.1234  
  \end{array}
  $$
  
  \vskip 2cm
  
  Table 2. Parameters and characteristics of regular solutions for ansatz  $\bf B$. 
  $$                     
  \begin{array}{lccccccc}
  n & \beta=k(0) & \delta=U(0) & \varepsilon=U(\infty) & W_n  &  W_n^{var} & W_n\cdot n^2 & W_n^{var}\cdot n^2\\ [7pt]
  2 & 0.020682 & 0.07992 & 0.001025  & 0.02664 & 0.02534 & 0.1066 & 0.1014 \\
  3 & 0.013741 &  0.03797 & 0.000705 & 0.01266 & 0.01272 & 0.1139 & 0.1145 \\
  4 & 0.009477 & 0.02178  & 0.000400 &  0.007256 & 0.007407 & 0.1161 & 0.1185\\ 
  5 & 0.006960 & 0.001405   & 0.000253 & 0.004682  & 0.004813 & 0.1174 & 0.1203\\
  6 & 0.005366 & 0.009791 &  0.000168  & 0.003264 & 0.003369 &  0.1175 & 0.1213\\
  7 &  0.007205 &  0.004292 &  0.000116 & 0.002402 &  0.002487 & 0.1177 & 0.1219\\
  8 & 0.003530 & 0.005524 & 0.0000861 &  0.001842 & 0.001910 & 0.1179 & 0.1222\\ 
  9 & 0.002966 & 0.004367 & 0.0000618 & 0.001455 & 0.001512 & 0.1179 & 0.1225\\
  10 & 0.002540 & 0.003539   & 0.0000492 & 0.001180 & 0.001227 & 0.1179 & 0.1227\\ 
  11 & 0.002205 & 0.002925 & 0.0000379 & 0.0009751 & 0.001015 & 0.1180  & 0.1228 \\ 
  12 & 0.001938 & 0.002458  & 0.0000285 & 0.0008194 & 0.0008535 &  0.1180 & 0.1229\\
  ... & ... & ... & ... & ... &  ... & ....& .... \\[7pt]
  52 & ..... & .......& ...... & ...... & 4.563 \cdot 10^{-5} & .... & 0.1234 \\
  102 & ..... & .......& ...... & ...... & 1.186 \cdot 10^{-5} & .... & 0.1234  
  \end{array}
  $$

  \clearpage
\section{DISCUSSION OF RESULTS}

It was already noted that consideration of the electric self-field of the electron in the hydrogen atom
looks quite justified, if not inevitable, both from the
general physical point of view, and according to the
QED ideas. The mathematical structure of the system
of Dirac-Maxwell equations becomes quite different
from the canonical one; this system is now coupled and effectively nonlinear. 

The spectrum of regular states of such a system can
be rather reliably determined by numerical and variational methods only for the “nonlinear analogs” of  s- or p-states. Indeed, only in these cases successive
terms in the chain of equations for the radial functions
have increasing order of smallness with respect to the
fine structure constant. In the nonrelativistic approximation  $\alpha \rightarrow 0$ only the principal harmonics corresponding to the orbital quantum number $l=0,1$, contribute to the binding energy. It is possible to obtain
relativistic corrections determined for the same solutions by harmonics with higher values of $l$ and corresponding harmonics of electric and magnetic potentials of the electron;  however, it requires cumbersome
calculations and is not considered here (estimations
were obtained in~\cite{Hydrogen}). 

Thus, the solutions and the binding energy spectrum given in this paper should actually be compared
with solutions to the Schrödinger equation of s- and
p-types in the presence of the external Coulomb field. Surprisingly, quite a different structure of the considered
nonlinear model reproduces the canonical Bohrian
distribution with good accuracy, and the better, the
larger “principal quantum number” is considered.
The effective Rydberg constant turns out to have very
close values for both classes of solutions, thus, it can be
assumed universal. The spinor “wave function” normalization condition is satisfied as a direct consequence of the natural electroneutrality condition, and
the (projection) of the total angular momentum turns
out exactly equal to  $\pm \hbar/2$.

 Unfortunately, at this point coincidences with quantum
 mechanics and experiment cease. The numerical
 value of the “Rydberg constant” is approximately
 twice as small as the observed one, and the “Lamb
 shift” is larger than the observed one by many orders
 of magnitude. Also, the magnetic moment of distributions corresponding to the “analogs” of p-states,
 strangely, is three times smaller than the magnetic
 moment of the electron.
 It is possible, of course, following ~\cite{Pestov} to transform
 the discovered spectrum of levels using the classical analog of the renormalization procedure that would have
 nothing to do with subtracting of infinities, like in QED,
 but would be reduced to simply equating the observed
 electron mass to the doubled initial mass of the bare particle $m$ in the Dirac equation. However, this trick, obviously, does not solve all problems of agreement between the obtained results and experiment, including the value of magnetic
 moment of the ansatz {\bf B}  and the anomalously large difference between the energies of levels $2s_{1/2}$ and $2p_{1/2}$.          
 
 In general, it seems that complete agreement with
 experiment, including radiative corrections, is possible in the framework of a purely classical field model of
 the hydrogen atom free of any divergences and the use
 of second quantization. For this purpose, however, it
 is necessary to consider a certain additional factor, for
 example, self-action with respect to the spinor field~\cite{Soler}  or possible supersymmetric structure of the considered self-consistent model~\cite{Wipf}. We leave more
 detailed discussion of these issues for later. 
 
\section{ADDENDUM}

In the reviewing process, a paper by Ranãda and
Uson ~\cite{Ranada} was found, which is very close both with
respect to the problem formulation, methods of study,
and results obtained~\footnote{We are deeply grateful to the referee for attentiveness, important remarks and sending us paper ~\cite{Ranada}.}. Namely, in essence, a similar hydrogen
atom model accounting for the electric self-field of the
electron represented by the self-consistent system of
Dirac-Maxwell equation with Lagrangian (\ref{Lagrang}) was
considered. The same stationary regular solutions
related to the nonlinear analogs of s- and p-states and
corresponding to our ansatzes {\bf A} and {\bf B}, respectively,
were sought.

However, instead of direct reduction to the nonrelativistic case based on transformation (\ref{bor})  and subsequent $\alpha$-expansion (Section 3), a cumbersome procedure of the numerical integration of initial system  (\ref{eqsstat}) (in the unsubstantiated assumption on the smallness
of terms with magnetic potential in the equations for
spinor harmonics) was applied in~\cite{Ranada}. 

Nonetheless, for the first three levels~1s, 2s, 3s and
two levels~ 1p, 2p  the results obtained in ~\cite{Ranada} agree with
those given above (Tables 1 and 2) with good accuracy.
For example, for the energy $E$ of the ground state 1s,
according to formula (26) from~\cite{Ranada}, one finds, up to
corrections $\sim \alpha^4$):   
\be{boundenergy}
E= mc^2 (1 - 0.243894 \alpha^2),  
\ee
which, similar to our study, corresponds to the binding
energy $W_B \approx R/2$ and is approximately twice as small
as the experimental value. For the first three s-levels
they have, with good accuracy, Bohrian function for
the binding energy (\ref{numbohr}) (our result is first 10 levels in
the case of the numerical method, and up to $n=100$ in the case of the variational method). The results for
the first two p-states considered in~\cite{Ranada}, also agree well
with the above data.

Correspondingly, the conclusion made in~\cite{Ranada} is similar to
ours. Namely, in spite of the physically grounded
statement of the problem, the results do not improve but violate the agreement with the linear quantum mechanical problem and experiment.
In this relation, similar to ~\cite{Pestov}, it is proposed in~\cite{Ranada} to apply the classical analog of the renormalization
procedure by a corresponding redefinition of the coupling constant $e$ and the “bare” mass $m$, in order to formally agree with the observed parameters. In this case,
it is possible to provide correct values for the energy
difference and absolute values for the first three s-levels. According to the authors, for p-states the results of reconciliation are not quite satisfactory.

Unfortunately, as it was already noted above (Section 6), in the considered model, the charge/mass renormalization is in principle incapable of solving the
problem of achieving agreement with experiment.
Obviously, say, it cannot change the ratio of magnetic
moments of the particle in s- and p-state $M_s/M_p = 3$, or the anomalously large difference between the binding energies for $2s_{1/2}$ and $2p_{1/2}$ states. Note that in~\cite{Ranada} these effects were not discovered at all. 

Moreover, the
absolute values of the mechanical $J$ and magnetic $M$ moments of the particle change in the case of renormalization proposed in~\cite{Ranada}. For example, for s-states
we have 
\be{renorm}
J = X J_0, M = X M_0, ~~ J_0: = \frac{\hbar}{2},~M_0 = \frac{e_0 \hbar}{2m_0 c},   
~~X:=\frac{e_0}{e}\equiv\frac{m_0}{m}, 
\ee
where $e_0, m_0$ are the electron charge and mass, respectively, and  $e, m$  are the bare charge and mass, i.e., the corresponding scaling factors in initial Lagrangian  (\ref{Lagrang}). For formal agreement of the energy of s- and p-states
with experiment, the normalization $X$ should be taken
as follows\footnote{According to our calculations (compare with (\ref{dimbohr})) a more accurate value is as follows: $X^2\approx 1/0.48\approx 2.078$,  and according to
	the data for the first three s-levels in~\cite{Ranada}, $X^2\approx 2.0453$}: $X^2 \approx 2$.  This scale transformation results thus 
in incorrect values of the mechanical and magnetic
moments, which was not noticed in~\cite{Ranada}

Finally, charge/mass renormalization violates the
normalization condition of the particle spinor field (\ref{Neutral}), which prevents us from considering it as the “wave
function”; it also violates the agreement of the model
with the linear quantum theory on the whole. Summing up the above said, it should be admitted that in
the form considered here and in~\cite{Ranada}  the Dirac-Maxwell model, being mathematically correct and
extremely rich with respect to its inner properties, is
unsatisfactory from the physical point of view and
should be substantially changed.
Note in this regard, that in ~\cite{Ranada} an important conclusion on a negligibly small effect of self-action with
respect to the spinor field (proposed above, see Section 6) on the hydrogen spectrum was made. The grounds for this conclusion are both the results of the
numerical integration and qualitative considerations
on large spatial extension of the Dirac field ($> 10^{-8}$ cm) in the hydrogen atom, i.e., extension to the region
where the nonlinear term is negligible. Thus, other
possibilities for the model modification should be
used. We expect to discuss and consider some of them
in the future.

\section{ACKNOWLEDGMENTS}
We thank A.B. Pestov, Yu.P. Rybakov, and B.N. Frolov
for helpful discussions and bibliographic indications.

\section{FUNDING}
The work was supported by the “University program
5-100” of the Russian Peoples’ Friendship University.      
	\selectlanguage{English}

\end{document}